\documentclass[preprint,12pt,authoryear]{elsarticle}
\usepackage{xcolor}
\usepackage{csquotes}

\usepackage{amssymb}

\usepackage{hyperref}
\usepackage{caption}
\usepackage{subcaption}
\usepackage{lscape}

\usepackage{tikz}
\usetikzlibrary{shapes}
\usepackage{float}

\newcommand{\D}{8} 
\newcommand{\U}{10} 

\newdimen\R 
\newdimen\L 
\newcommand{\A}{360/\D} 

\journal{International Journal of Educational Research}

\begin{document}

\begin{frontmatter}



\title{A Scheduling Perspective on Modular Educational Systems in Europe}


\author[imc]{Rub\'en Ruiz-Torrubiano\corref{cor1}}
\ead{ruben.ruiz@fh-krems.ac.at}

\author[untis]{Sebastian Knopp}
\ead{sebastian.knopp@untis.at}

\author[untis]{Lukas Matthias Wolf}
\ead{lukas.m.wolf@untis.at}

\author[imc]{Andreas Krystallidis}
\ead{andreas.krystallidis@fh-krems.ac.at}

\affiliation[imc]{organization={IMC Krems University of Applied Sciences},
            addressline={Piaristengasse 1}, 
            city={Krems},
            postcode={3500}, 
            country={Austria}}

\affiliation[untis]{organization={Untis GmbH},
            addressline={Belvederegasse 11}, 
            city={Stockerau},
            postcode={2000}, 
            country={Austria}}

\cortext[cor1]{Corresponding author.}

\begin{abstract}
In modular educational systems, students are allowed to choose a part of their own curriculum themselves.
This is typically done in the final class levels which lead to maturity for university access.
The rationale behind letting students choose their courses themselves is to enhance 
self-responsibility, improve student motivation, and allow a focus on specific areas of interest.
A central instrument for bringing these systems to fruition is the timetable.
However, scheduling the timetable in such systems can be an extremely challenging and time-consuming task.
In this study, we present a framework for classifying  
modular educational systems in Europe that reflects different degrees of freedom regarding student choices,
and explore the consequences from the perspective of scheduling a timetable that satisfies all requirements
from the organizational and the pedagogical perspective. For this purpose, we conducted interviews in Austria,
Germany, Finland, Switzerland, the Netherlands, and Luxembourg and apply the framework to these educational systems,
finding that among them the Finnish system shows the highest degree of modularity. After 
analyzing the consequences of modularity from the scheduling perspective, we assess the necessity for automated scheduling methods,
which are central for realizing the potential and many benefits of modular education in practice.


\end{abstract}


\begin{highlights}
\item We present a category framework to classify educational systems according to their degree of modularity.
\item A graphical comparison is given that shows how different educational systems score on modularity according to our framework.
\item A scheduling perspective is provided where the challenges regarding building timetables for modular educational systems are investigated.
\item To the best of our knowledge, this is the first study of this kind to date.
\end{highlights}


\end{frontmatter}


\section{Introduction}\label{introduction}

Curriculum design is one of the central activities of stakeholders involved in educational systems 
in order to steer and implement strategic goals that education should fulfill in a given country or region. 
One of these goals can be considered in general to meet the current needs and demands of society
and the expectations of the population, especially students, parents and teachers.
Depending on the stakeholders and the output of this activity, different levels of curriculum 
design can be defined \citep{van_den_akker_curriculum_2007}: international/comparative (\emph{supra}), 
system/nation/state (\emph{macro}), school or institution (\emph{meso}), classroom (\emph{micro}) and
individual level (\emph{nano}). The international or supra level is usually associated with comparative studies 
(e.g., PISA) and reflects on aims and quality of education in general. The macro level is focused on the 
particular implementation of those goals in a given societal context. This is normally achieved in the
form of national syllabi or concrete objectives. The meso level is concerned with the implementation 
of the curriculum in the context of a particular school. For instance, a school might decide to offer 
additional electives or make decisions on how parts of the curriculum are delivered. Finally, the micro and 
nano levels are mostly concerned with the classroom implementation of the curriculum and how it is 
handled by teachers and students.

In particular, it is important to consider how to enhance student engagement (i.e., the nano level) 
already at the macro level. In practice, this means laying the regulatory groundwork for
improving student involvement and particupation at the lower levels. One of the most effective 
approaches in this context is the introduction of electives in the curriculum, which is often
referred to as modular course design. The origins of modularization in education can be traced 
back to Harvard University the 19th century \citep{bell_modular_1993}, where an elective system was introduced to increase 
the choices available for students and increase the flexibility of the curriculum. This development 
was explicitly encouraged in Britain in the 1960s and 1970s for increasing participation rates and 
enhance teacher and student mobility. Moreover, modular education has been shown to 
significantly decrease school dropout rates \citep{mazrekaj_effect_2020} and improve student
achievement \citep{vidal_rodeiro_effects_2012}.

When implementing curricula at the meso (school) level, one of the most challenging operative and 
organizational questions is how to schedule courses in a timetable, possibly including both modular 
and non-modular components. How courses are scheduled can affect the academic performance of students 
\citep{dills_course_2008}. From the organizational perspective, the timetable can be considered 
as the central school resource plan that coordinates how educational resources (teachers, rooms, 
laboratories, etc) are used. Constraints arising from a variety of sources, including teacher and 
room availability, students timetable quality and pedagogical considerations can make timetabling very
challenging in practice.

In this study, we aim to answer the following research question: \emph{How do different educational systems 
in Europe implement modular education and how this relates to timetable scheduling from an 
organizational and pedagogical perspective?"}. We note that the focus of our work is placed on 
secondary rather than tertiary eduction. To answer the research question, a novel theoretical framework
is developed to assess the degree of modularity in those systems in a qualitative and quantitative way. 
We focus on the educational systems of Austria, Germany, Finland, the Netherlands, Switzerland and 
Luxembourg and perform semi-structured interviews with experts and practitioners to gather insights 
into how modularity influences
scheduling and what are the most important problems in practice. This framework is the 
first step in a larger effort to propose efficient computational methods for solving the scheduling 
problem in modular educational systems in practice. To the best of our knowledge, this is 
the first study that investigates modular education from a scheduling perspective.

This article is organized as follows. In Section~\ref{methodology} we outline the methodology used for 
this study and introduce central definitions used throughout the paper. Section~\ref{results} presents 
the results of the interviews and the proposed framework. A detailed discussion of our findings is 
provided in Section~\ref{discussion} and we conclude in Section~\ref{conclusion} with a summary and 
future work. 

\section{Methodology}\label{methodology}

\subsection{General methodology}

We perform a qualitative content analysis \citep{kohlbacher_use_2006} based on Mayring's
approach \citep{mayring_qualitative_2003}. The main difference between this approach and standard
content analysis is that quantitative analysis (e.g., frequency analysis) is not used as a means for 
evaluating the resulting categories. Instead a systematic, theory-guided approach is used for 
text-analysis that takes the context into account. Therefore, the focus of the analysis is on
the qualities, the processes and the meanings rather than on experimentally measurable entities.
Furthermore, Mayring's approach is theory-guided, which allows the integration of previous
knowledge and theoretical frameworks into the analytic process. 

Our study can be framed as an instance of case study research. The goal of case study research 
is to investigate phenomena within their context \citep{hartley_case_2004}. Case study as a 
research strategy is particularly well suited when the subject of the investigation includes 
organizational or managerial processes \citep{yin_case_2003}, as in the current context. Our focus 
is on the organizational processes dedicated to scheduling in the context of the particular 
modular educational system under consideration. Therefore, we conducted semi-structured interviews 
with 12 experts with direct experience in scheduling timetables in modular 
educational systems. The interviews were performed from January until July 2023.
Additionally, interview notes and internal documents were also incorporated into the analysis. 
After conducting the interviews, an inductive approach was used to develop and refine 
the categories. Coding was performed in a collaborative way by iteratively achieving mutual consent
regarding the results of the analysis in what is known 
as communicative validation \citep{mayring_qualitative_2003}.

\subsection{Interview partners and regions}

In Table~\ref{table_interviews} we provide a summary of all the interview partners involved, including 
a short description of their expertise and the country/region they work in. As can be seen from 
the table, seven of the experts are timetablers and therefore with relevant practical expertise 
in the organizational processes of the school. Additionally, six interview partners are experts in 
timetabling software with extensive knowledge of their respective modular education system. 
Finally, one of the interview partners works at the local Ministry of Education and contributes with 
extensive knowledge on the educational system. Note that in total eight out of the total of twelve experts 
come from a German-speaking country, and several German regions are represented separately. The reason 
is that the authors already have experience with these regions. Additionally, the federal nature 
of the German educational system makes it necessary to investigate regions separately, as 
differences in curricula and school types can be significant.

\begin{table}[h!]
\small
\centering
\begin{tabular}{|l|l|l|l|}
\hline
ID & Country (region) & Description & Experience \\
\hline \hline
E1 & Austria & High school timetabler & More than 5 years \\ 
 & (Lower Austria) & (\emph{Gymnasium}) &  \\ \hline 
E2 & Germany & Timetabling-software expert & More than 25 years \\ 
 & (Baden-Wurttemberg) & &  \\ \hline
E3 & Germany & High school timetabler and & More than 25 years \\ 
 & (Baden-Wurttemberg) & timetabling-software expert &  \\ \hline
E4 & Germany & High school timetabler and & More than 25 years \\ 
 & (Bavaria) & timetabling-software expert &  \\ \hline
E5 & Germany & Timetabling-software expert & More than 15 years \\ 
 & (Lower Saxony) & &  \\ \hline
E6 & Germany & High school timetabler & More than 25 years \\ 
 & (Rheinland-Palatinate) & &  \\ \hline
E7 & Germany & Modular education expert & More than 25 years \\ 
 & (Rheinland-Palatinate) & at local Ministry &  \\ \hline
E8 & Netherlands & High school timetabler & More than 25 years \\ 
 & & &  \\ \hline
E9 & Luxembourg & High school timetabler & More than 5 years \\ 
 &  & (\emph{European school}) &  \\ \hline 
E10 & Luxembourg & High school timetabler & More than 10 years \\ 
 &  & (\emph{European school}) &  \\ \hline
E11 & Switzerland & Timetabling-software expert & More than 25 years \\ 
 & (German-speaking) & &  \\ \hline
E12 & Switzerland & Timetabling-software expert & More than 10 years \\ 
 & (French-speaking) & &  \\ \hline
 \hline
\end{tabular}
\caption{Summary of interview partners.}\label{table_interviews}
\end{table}

\subsection{Coding of the interviews}

Once conducted, the interviews were transcripted and annotated. Since the authors have 
extensive experience with different educational systems, this a-priori knowledge was 
incorporated in the coding process as a set of pre-defined codes. These codes included 
terms like "election", "module", "scheduling" and "choices". Once the interviews were 
encoded using these terms, the codes were refined to take recurrent themes like 
"election systems", "class level overlaps", "number of choices", etc, into account. The refined codes 
were obtained inductively by finding commonalities and recurrent themes between the contents of the different 
interviews. For instance, it was a common theme to describe what the system of electives 
looked like in terms of number of electives or student choices made.

After going through several iterations of this method, consensus was reached on the final categories,
including a categorization of each individual educational system into this set of categories. In 
Section~\ref{results}, we present the result of this categorization in detail.

\subsection{Basic definitions}

In this section, we define some basic terms that will be used later on
to classify the modular education systems that we study in this paper.

\begin{description}
    \item[Class] A group of students that attend all regular lessons together. By 
        regular lessons we mean those that are mandatory for all students (i.e., they cannot be 
        elected).
    \item[Class level/grade] The yearly part of a class, e.g., the fourth grade, the fifth grade, etc.
        Typically, there are several classes belonging to a given class level (i.e., the classes 
        4a, 4b, etc.). In what follows we will refer to class levels meaning groups of classes with the same yearly part.
    \item[Lesson] A lesson is an event where a group of students belonging to one or several classes are taught in a given subject. Usually 
        the total regular weekly duration of a lesson is split into sessions that can have different lengths. 
        For instance, a math lesson can have a regular duration of 3 hours per week and be split 
        into one two-hour and one one-hour session. Sessions are normally constrained by standard times defined 
        in a \emph{time grid}. For instance, a school might define that each day lessons start at 8.00 am 
        and each session has a duration of 50 minutes. In this case, instead of talking about hours we will normally talk 
        about \emph{time slots} or \emph{periods} (both terms are interchangeable).
    \item[Courses/Electable lessons]
        Lessons are called electable if they are not mandatory for a student, i.e., they can be selected
        eventually subject to some restrictions (e.g., pre-conditions like previous knowledge/attended courses).
        In this paper, we will use the term course to mean an electable lesson. 
    \item[Student election/choice] We will refer to the action of a student selecting a course/module
        as an election or choice (these terms can be used interchangeably). 
    \item[Term] This is a subdivision of a school year into disjoint intervals. Most typically, 
        a term can represent a semester (e.g., fall/spring semesters), but other, shorter terms are also possible.
\end{description}

\section{Results}\label{results}

We now describe the category system resulting from encoding the case-study interviews 
presented in Section~\ref{methodology}. First we describe the category system, which is composed 
of a total of eight categories. Then, we apply these categories to the individual educational systems 
that were investigated in the conducted interviews by evaluating each category in the specific 
context of each system. In Section \ref{discussion}, a visual comparison between the 
investigated educational systems regarding their degree of modularity is provided along with a detailed discussion of the results.

\subsection{Categories}

After conducting the qualitative content analysis and coding process described in 
Section~\ref{methodology}, the following categories were identified: 

\begin{description}
    \item[Proportion of electable periods (PEP)]
        Each modular educational system investigated in this study shows varying degrees regarding the 
        proportion of electable lessons. Given a class level, we define the proportion of electable periods in 
        that class level as the total number of periods of courses that can be elected by students in that class level
        divided by the total number of periods of all lessons (electable and mandatory). So 
        for example, if there are 5 lessons in class level 8,
        and 2 of the 5 lessons are courses, and all lessons have a duration of 2 periods,
        the PEP of class level 8 would be $(2 \cdot 2)/(5 \cdot 2)=0.4$, or 40\%. We take the maximum of all 
        class levels as the PEP of a given school form. E.g., if in one class level everything can be chosen freely,
        we consider this as 100\%, even if in lower class levels nothing can be chosen.
    \item[Election systems (ES)]
        In general, there are different ways of offering electable lessons to students. After conducting the 
        interviews, we distinguish between three different module selection systems:
        \begin{description}
            \item[Class-based]
                Each student is assigned only to lessons that are planed for the class the student belongs to.
            \item[Profile-based]
                Each student can choose one out of a list of pre-defined \emph{profiles}. A profile determines a set of 
                courses which all students who have chosen the same profile take together. For example, it might be 
                possible in a given school to choose between \emph{natural sciences} and \emph{humanities}. The profile 
                \emph{natural sciences} could include the courses mathematics, physics and chemistry and \emph{humanities}
                could include greek, latin and philosophy. All the other lessons are taught class-based (i.e. are mandatory). 
            \item[Individual]
                Each student can choose among a pool of courses individually with a given level of freedom. Consequently, each 
                course can be composed of a different group of students that might be unique for that course. Note that 
                in individual election systems it might also be the case that some proportion of the lessons are taken 
                class-based (i.e., core subjects like mathematics and history).
        \end{description}
        As will be shown later, it is also possible to have profile-based systems where students can choose additional 
        subjects outside of their profile (e.g. in the Netherlands). We consider theses systems to be halfway between a profile-based and a pure 
        individual election system.
    \item[Number of student elections (SE)]
        During their stay at school, students in modular systems will need to make
        a number of elections for courses taught in a given class level (e.g. by filling some forms).
        For example, if students are asked to select courses only in each of the last three years,
        we define the number of student elections to be three. 
    \item[Proportion of module-oriented class levels (ML)]
        This is the proportion of class levels offering either profile-based or individual course elections. For example, 
        if there are 5 class levels and 3 of them offer electable lessons, this proportion would be $3/5=0.6$ or 60\%.
    \item[Number of terms per year (T)]
        The number of different periods of time characterized by different regular timetables. In systems where there is a 
        fall and a spring semester, the number of terms per year would be~2. In other systems (e.g. Finland) there might 
        be an arbitrary number of terms higher than 2.
    \item[Scheduling workflow (SW)]
        We identified two different scheduling workflows from the experts for creating timeables in modular systems.
        Either the timetable is created first and students choose courses based on the known timetable
        or students select courses based only on their preferences without knowing the timetable in advance.
        In this last case, the timetable is created taking all student elections into consideration. We call the 
        first workflow the \emph{timetable-first} and the second the \emph{election-first} workflow. 
    \item[Proportion of courses with class level overlap (CLO)] 
        We say that a course shows \emph{class level overlap} if students from different class levels are allowed to attend 
        the course (i.e., the course is electable for students in different class levels). For instance, a philosophy course 
        might be offered as electable for all students in the seventh and the eighth grade, which means that potentially 
        students from both levels can be taught that course simultaneously.
    \item[School course collaborations (SC)]
        In some systems, rare or unusual subjects requiring teachers with special qualifications might be offered 
        collectively by two or more schools to reduce costs and optimize attendance. In this case, guest students from 
        other schools are allowed to attend a course given in a specific host school. 
\end{description}

Table~\ref{table_categories} presents a general overview of our category system. In addition to the 
name of the category, the abbreviation, and a short description, the last column specifies which range of values 
each category is allowed to take. For proportions, in general we divide the range of possible values in three 
intervals \emph{low}, \emph{medium} and \emph{high} that correspond each to one third of the $[0, 100]$ interval.
Cardinal values are positive integers. All the other categories take values in a definite set (e.g., "Yes/No"). In the 
following, we review the secondary modular educational systems of Austria, Germany (for selected federal states),
Switzerland, Finland, Luxembourg and the Netherlands.

\begin{table}[h!]
    \small
    \centering
    \begin{tabular}{|l|l|l|l|}
    \hline
    Category Name & Abbr. & Description & Values \\
    \hline \hline
    Prop. electable periods & PEP & How many periods are & Low, medium,  \\
    & & electable?  & high \\
     \hline
     Election systems & ES & Which framework is used for & Class, profile, \\
    & & electable courses?  & individual \\
    \hline
    No. student elections & SE & How often have students to & Low, medium, \\
    & & select courses?  & high \\
    \hline
    Module-oriented class levels & ML & How many class levels & Low-medium, \\
    & & contain modular courses?  & high \\
    \hline
    No. terms/year & T & How many different terms  & Cardinal \\
    & & take place in a school year? & \\
    \hline
    Scheduling workflow & SW & How is the timetable  & Timetable-first, \\
    & & scheduled? & Election-first \\
    \hline
    Prop. courses with class & CLO & How many courses are mixed  & Low, medium \\
    overlap & & among class levels? & high \\
    \hline
    School collaborations & SC & Are courses offered among   & Yes/No \\
    & & different schools? &  \\
    \hline
    \end{tabular}
    \caption{Summary of the resulting categories.}\label{table_categories}
\end{table}


\subsection{Austria}

In Austria, secondary education is distributed among different school types \citep{noauthor_schulorganisationsgesetz_2001}. The type of high school 
known as \emph{Gymnasium} is the most relevant from the point of view of modular education. In general,
class levels are grouped into two cycles of four years each, the lower cycle (or \emph{Unterstufe}, US) from the 
first until the fouth grade and the upper cycle (or \emph{Oberstufe}, OS) from the fifth until the eigth grade. From 
the sixth until the eighth grade there is a minimum of four and a maximum of ten hours per week of 
electives that can be offered by each school autonomously. Modular education is still under development in Austria. 
One of the earliest forms of modularization in secondary education 
was the MOST (modular uppper cycle), which has been tried individually at some schools since the late 2000s. This school 
form was unified in what was called NOVI (new upper cycle with strengthened individualization), which is being tested at 
the time of this writing in about 30 schools. This model is planned to evolve into the SOST (upper cycle in semester-based form), that 
changes the modules' assessment from a yearly to a semester basis \citep{oead_what_nodate}. 

In the NOVI, there are up to eleven elective modules that can be chosen without any pre-conditions. These eleven courses are 
typically taught for two hours per week and distributed evenly
among the last two class levels in the OS (seventh and eight). That means that, in a typical week, students attend about 30 hours of 
non-electives and 4 hours in modules. To facilitate scheduling, all non-electives are planned in the morning, the modules being taught 
in the afternoon.

\paragraph{Category classification}
Accoding to our category system, the proportion of electable periods can be considered to be on the lower end of the scale.
Using the typical scenario outlined before, this would amount to about one third of the total number of periods per week to be in 
modules. The module election sytem is individual: there are no restrictions for students on which modules they are allowed to select.
The number of choices that a student has to make lies within the lower end of the scale (students only choose their subjects once 
and this remains until the end of the upper cycle). In this system, only the last two class levels are module-oriented, i.e., from 
the total of eight levels that would result in a total proportion of 25\%. In a typical year, there are two semesters (fall/spring).
Regarding scheduling workflow, first the timetable is made, and after that the students can select their subjects based on the 
already existing schedule. Class level overlaps are usual (students from the seventh and the eighth grade can be present) and 
there are no school collaborations.

\subsection{Finland}

Finland has a long tradition in modular education at the upper secondary level (high school), comprising the last years levels where 
students are generally between 16 and 18 years old.
High schools therefore normally encompass three years although some students have to take a fourth. All students participate in the elective 
course system. Courses are not necessarily bound to class levels and therefore most courses can be elected by students in any of them.
We see around 75\% elective courses in course election data, the others being mandatory.

\paragraph{Category classification}
The choice system is individual, except for general class hours with a class teacher in one subject. The number of choices is as follows:
Students in the first two class levels choose between 30-40 courses, while students in the third and last year can select between 20-24 courses.
There are between 5-6 terms per year and courses can take place in any of them, 
but there is also the possibility for students to take a certain exam so the corresponding course 
has to take place in the corresponding term too.
Students first make their course choices, afterwards the timetable is created (election-first workflow).
Usually, the assignment of teachers to courses is not known at the time the student make their choices. School collaborations are 
usual, especially in the context of a given municipality. High schools in the same municipality share courses that students of 
other schools can attend. This results in a very high complexity when scheduling courses, since times for shared courses need 
to be taken into account.

\subsection{Germany}

In Germany, the responsibility for the school system lies with the 16 federal states.
Consequently, significant differences between them exist.
As in Austria, secondary education is distributed among different school types.
Regarding modular education, school types which lead to the
general qualification for university entrance (\emph{Abitur}) are the most relevant.
These are of the types \emph{Gymnasium}, \emph{Oberschule}, or \emph{Gesamtschule}.
In particular in the two or three highest class levels (upper cycle)
electives can be chosen by students with a high degree of freedom.

In the following subsections,
we present the situations in specific federal states.
We conducted five interviews with seven experts for course timetabling.
These interviews cover schools in 10 of the 16 federal states of Germany:
Two experts for Baden-Württemberg, one expert for Bavaria,
three experts for Rhineland-Palatinate,
and one expert covering seven states
(Brandenburg, Bremen, Lower Saxony, Mecklenburg-Vorpommern, Saxony, Saxony-Anhalt, Thuringia)
at the same time.

\subsubsection{Baden-Württemberg}

In Baden-Württemberg, there are two different school types in secondary education.
One is the \emph{Allgemeinbildendes Gymnasium} which has an elective system
that allows many individual electives for students.
The other is the \emph{Berufliches Gymnasium} where
more periods per week (about 12 to 14) take place within a class context
and a profile based election system is used.
In both school types, a course-based system is used in the final two year levels.
In Baden-Württemberg, there are either 12 or 13 years levels for each student since primary school.
So, depending on this, the course-based system is used
in year levels eleven and twelve, or twelve and thirteen.

\paragraph{Category classification}
As \emph{Allgemeinbildende Gymnasien} (AGY) and \emph{Berufliche Gymnasien} (BGY)
are quite different, we classify them in separate categories.
The proportion of electable periods is medium (BGY) or high (AGY).
All periods take place in modules (AGY) or roughly half of the periods (BGY).
The module election sytem is individual (AGY) or profile based (BGY).
The remaining category classification is identical for both school types.
Students make their choices once, right before they enter the class levels of the upper cycle.
So, the number of elections is low.
In this system, only the last two class levels are module-oriented, i.e.,
from the total of eight (or nine) levels that would result in a total proportion of~25\% (low).
Timetables can be different in each term (half-year).
An election-first scheduling workflow is used.
Students first make their course choices, after that the timetable is created.
The assignment of teachers to courses is not known at the time the student make their choices.
Class level overlaps can be planned for lessons which are chosen by only few students.
This happens rarely (low).
For the same reason, some schools in this state do collaborate to offer common courses.

\subsubsection{Bavaria}

In the federal state of Bavaria, modular education is mainly present at two school types: the 
\emph{Fachoberschule} (FOS) and the \emph{Berufsoberschule} (BOS). The main difference lies in the 
practical orientation: while the FOS can be attended by all students with a middle school diploma,
the BOS is intended for students with some professional experience that aim at deepening their 
training in that area \citep{noauthor_schulordnung_bayern_2017}. 
Both types of schools lead to the general qualification for university entrance. The FOS comprises 
three class levels (11, 12 and 13), while the BOS encompasses only two (12 and 13). Students choose
electives in the eleventh class level, having more choices in the BOS than the FOS. In general,
besides general subjects like math, history and ethics, students can choose one out of seven different 
profiles, each with their own profile subjects (e.g. environmental sciences, health sciences, 
technical subjects, etc.). Furthermore, and based on the chosen profile, students can choose up to two 
additional elective courses that aim at deepening the knowledge of the student in a particular direction
(for instance, diverse language courses, biotechnology, arts and music, etc.). In total, around 20 hours
per week are dedicated to general subjects, 10 for profile subjects and 4 hours for freely electable 
modules. 

\paragraph{Category classification}
The proportion of electable periods can be considered medium (about 50\% of the time is spent in 
either profile or electable courses). Therefore, the module election system is a mixture of profile-based and 
individual system. Students typically choose once their profile and electives (in the eleventh class level in the FOS or during 
their practical training in the BOS), and these elections remain constant until their graduation. All class 
levels except the eleventh (FOS) are modular, and there are in general no distinction between semesters. 
The scheduling workflow is election-first: students choose their profiles and electives first, and then the 
timetable is made with the goal of fulfilling all the student elections. Class level overlaps are possible, 
and in general there are no school collaborations. 

\subsubsection{Rhineland-Palatinate}

Modular education in Rhineland-Palatinate is mainly applied in the last three years of the \emph{Gymnasium},
which can be either the 10th, 11th and 12th grades in 8-year schools (G8), or the 11th, 12th and 13th grade 
in 9-year schools (G9). Across all class levels, students take between 32 and 35 hours per week of courses, 
which are divided into basic courses (which are taught for 3 hours per week) and advanced level courses
(which are taught 4 or 5 hours per week). In general, a student will choose 3 advanced courses,
the rest being basic courses on subjects not considered central regarding the profile of the student or the
school itself. In practice, resources are bundled together in some cases by defining advanced courses as 
basic courses with additional hours where more advanced material is taught. 

The course election process is normally coordinated by a single person, whereas another person is in charge 
of scheduling the courses in the timetable. On each year, students are asked to elect courses until March, and 
then groups of courses that can be scheduled together are clustered. Later on, the timetable is made 
according to those clusters. If possible without creating any disadvantage for other students, elected courses can be 
changed until the next autumm. 

\paragraph{Category classification} 
The \emph{Gymnasium} upper cycle in Rhineland-Palatinate can be considered as a medium-level modular system based 
on individual student elections. The core of this system is given by the election of the 
advanced courses. Since those courses are taught for either 4 or 5 hours per week, that results in between 12 and 
20 hours per week spent on advanced courses, which roughtly amounts to half the weekly lesson load for a typical student.
Only class levels in the upper cycle are modular, resulting in about 1/3 of modular class levels in the Gymnasium, 
and also here there is no distinction between semesters. The
scheduling process here is, as typical in Germany, also election-first. Class level overlaps are usual, and 
some schools may collaborate with others by bundling educational resources.

\subsubsection{Brandenburg, Bremen, Lower Saxony, Mecklenburg-Vorpommern, Saxony, Saxony-Anhalt, Thuringia}

Given that one of our interview partners is an expert on the educational systems of these federal states, and since 
these are quite similar, we proceed to report on all of them in this section. In general, we can distinguish between 
profile-based (e.g. in Bremen) and individual-election systems (e.g. Lower Saxony). In the former case, a high portion of the courses 
students attend are fixed by their profiles, whereas in the latter, students choose their courses individually. 
Depending on the specific federal state, the Gymnasium is divided into 9 class leves (G9), mostly in Bremen and 
Lower Saxony, or 8 levels in the rest (G8). Similarly as in Rhineland-Palatinate, only typically the last three levels 
are modular. While in the 11th grade a typical student spents about half of their time in modular courses, the 12th 
and the 13th grade (where appropriate) is completely modular, so students are allowed to elect all of their courses. Typically,
students will choose in the 11th class their courses, and once again in the 12th. 

\paragraph{Category classification}
In this case, we can regard these federal states as having highly modular educational systems, since the average amount 
of hours spent on modular courses by the average student is clearly higher than 50\%. There is a mixture of profile-based and 
individual election systems, although individual systems are more common. Like in Rhineland-Palatinate, about a third of all 
class levels in the Gymnasium can be regarded as modular. The
scheduling process is election-first, with frequent class-level overlaps and the possibility of school collaborations (mostly 
when these schools are located in the same city).

\subsection{Luxembourg}

The secondary school system in Luxembourg starts with 12 year old students and is divided into the following variations: classical seconday education (ESC), 
general secondary education (ESG), European education, International English curriculum, British education and 
German-Luxembourgish education. The high school education cycle in all cases always comprises 
seven class levels \citep{luxemburg_ministry_of_education_luxembourgish_2023}. Since both interviewees E9 and E10 
work at European schools, we will focus on this variant in the following discussion. In the European system, 
the 7-year cycle (S1-S7) is itself subdivided into three sub-cycles S1-S3, S4-S5 and S6-S7. In general, 
the degree of modularity increases with each sub-cycle. As part of the individual educational approach of each 
school (school autonomy), schools may introduce additional subjects or classes tailored to the particular approach 
of that school.

\paragraph{Category classification}

The proportion of electable periods starts in the first sub-cycle S1-S3 with approximately 30\% of periods 
in modular courses (including language courses) and increases to 50\% in S4-S5 until about 90\% in the last two years. The election system 
used is based on individual choices (no profiles are given). Students make course selections a total 
of three times, one per sub-cycle. All seven class levels of the secondary education system can be considered 
as modular, therefore the proportion of modular class level is highest. There is no special distinction between 
different terms in the course of a regular school year. The scheduling workflow used is a election-first type of 
workflow were students select their courses first, and then the timetable is scheduled according to the students 
choices. In general, students from different sections and class levels can be mixed into individual courses, 
like languages and natural sciences. School collaborations are not usual, since some schools are already
divided into different sites and this would overly complicate the scheduling process.

\subsection{Netherlands}

The secondary education system in the Netherlands is divided into pre-vocational (VMBO), pre-university (VWO) and senior general 
secondary education (HAVO) \citep{ministerie_van_onderwijs_secondary_2014}. The VMBO is profile-oriented. 
Once students choose a profile (an occupational sector), a predefined set of 
courses is automatically assigned. In general, students need to attend between 3 and 4 mandatory courses that make up a total 
of about 15 hours peer week for a typical student. In the HAVO and the VWO, students attend a pre-defined curriculum for the first 
three years. In the upper cycle, this common part of the curriculum gets reduced to core subjects like Dutch, English and 
social studies. Students then choose one among four different profiles (technology, health, economics and culture). However, students 
are allowed to choose up to two additional subjects related to the profile, and one additional elective, which might be 
unrelated to the profile. Regarding the timetable, schools try to assign students to courses so that the students stay most 
of the time in the same group, and avoid mixing students from different class levels (which is usual in other countries like 
in Germany). On the one side, this facilitates scheduling the different courses to a high degree, and 
on the other side helps students keep their social bonds during the time they spend in secondary school.

\paragraph{Category classification}
The degree of modularity of the different secondary education branches is slightly different for VMBO, VWO and HAVO, but in 
general one can consider the system to be medium modular. In all types of secondary education there are core subjects and optional
subjects, which are always pre-determined by the choice of a profile. Therefore, secondary education in the Netherlands can be 
regarded as profile-based. The number of choices that the pupils make is either only once (in fully profile-based variants 
like VMBO) or up to twice (like in the VWO). All class levels contain electives to some degree, and in general there is no 
distinction between fall and spring semesters. The schedule is made in an election-first manner, and if possible no students from 
different class levels are mixed in individual courses. School collaborations are not usual, since secondary schools in the 
Netherlands tend to be already quite large. 

\subsection{Switzerland}

For Switzerland, we conducted two interviews: one for the French-speaking cantons, and one for the German-speaking cantons.
We present the results in separate sections, as there are considerable differences between both approaches.

\subsubsection{German-speaking cantons}

In the German-speaking cantons of Switzerland,
modular education is mainly present at community schools (\emph{Gemeindeschulen}).
Most of the schools in these cantons are of this type,
overall the number of such schools is around~1000.
Although Gymnasiums also exist in these cantons, they almost don't offer any modular education,
therfore we don't consider them further in this paper. In the community schools, and 
all schools that implement the compulsory part of the education system, 
modular education is concerned basically with the last class level (that is the third 
level of the lower secondary cycle). However, there is no general obligation for schools 
to offer elective courses \citep{edk_lehrplan_2024}.
When they do, choices are made by students in spring before the timetabling takes place in summer.
Students choose specific courses without specifying alternatives.
In general, there is no collaboration between schools to offer shared courses.
For the timetablers, it is very important that idle periods 
are avoided for students in their timetables.

\paragraph{Category classification}
The proportion of electable periods is highly variable, but in general, the system can be considered to be of 
low modularity.
The module election sytem is individual.
In general, students choose only once (at the last year of their compulsory education).
The proportion of module-oriented class levels is therefore in general limited to one class level out of nine.
Timetables can be different in each term (half-year).
An election-first scheduling workflow is used.
Since in general only one class level is involved, there is no class level overlap in the elective courses.
As mentioned above, in general there are no school collaborations for shared courses.

\subsubsection{French-speaking cantons}

There are in total eight French-speaking cantons in Switzerland. In general, secondary education is divided into a 
lower and an upper cycle. The lower cycle is common for all of Switzerland, and lets students 
choose among three predefined profiles. However, the upper cycle (which involves students 
of ages between 15 and 19 years) can be tailored to specific cantons reflecting the diversity of approaches. In our interview, 
we focused on the Geneva, Fribourg, and Jura cantons as they stands out for their high degree of modularization. Before entering the upper cycle,
students are asked to select a basic set of courses, including two languages and a main optional subject
(students can specify alternatives in case the elected course is full or not available for some other reason). Students might 
also request to attend a course at another school, which in general is possible if they stay in the same canton. The mandatory 
subjects (French, geography, history and mathematics) account for about 30\% of the total lesson time for a typical student.
Students are asked to choose again courses in their third year (in the upper cycle), but the scope of this electives is much 
more reduced than in the first election. Scheduling is done by first finding a good placement for all the courses and then assigning 
the students to courses. 

\paragraph{Category classification}
We can consider the upper secondary education system in the above mentioned cantons to be modular 
to a high degree (i.e. about 70\% of the lesson time for a student is spent in elective courses). Students choose subjects twice during 
the upper cycle, once at the very beginning and once again in the third year. While students are mostly free to choose their subjects,
it results in practice into a large number of profiles (i.e. possible combinations of electives). However, our interview partner 
confirmed that from about 250 students, the number of resulting profiles can be as high as 200 or above. Therefore, for the purpose 
of this study we consider this system to be individually oriented, rather than profile-oriented. The proportion of modular class levels 
is 100\%, since both the lower and the upper cycle are designed in a modular way. There is no distinction between terms and the 
scheduling workflow is timetable-first (seemingly for historical reasons). Students from different class levels can be mixed in 
principle, and collaborations between schools are possible (e.g. a student attending courses in another school of the same 
canton). 

\section{Discussion}\label{discussion}

As can be seen in the interviews, the approaches to modular secondary education are very heterogeneous. However, 
some common aspects can be identified. For example, all of the aforementioned educational systems use modular education as 
a means of fostering individualization in students curricula. On the one side, individually-oriented election systems 
(like Germany or Finland) give more weight to students to follow their interest and deepen their understanding of core 
subjects. These electives might be relevant in their career after leaving secondary education, but professional usefulness alone does not provide 
the main rationale for the system. On the other side, profile-based systems (like in the Netherlands) are more usually focused on vocational 
education, and profiles are crafted in such a way that students learn to know which professions might be relevant 
in their professional life. In between, there are many different approaches to modular education as represented for 
instance in the case of Austria. In Germany, there are plans to harmonize the modular education system in the 
Gymnasium upper cycle to focus more on the advanced courses (as interview partner E7 mentioned) rather than basic courses,
stressing the focus of the system as enabler to deepen students' understanding of core subjects. How successful these plans 
turn out to be depends on future developments. As can also be seen in the case of Switzerland, countries with a highly federal 
structure tend to fragment their educational systems, making harmonization difficult in the long run.

Figure \ref{fig_comparison} shows graphically the results of the classification of all educational systems investigated 
in this work. For this purpose we use radar charts, which depict the different categories as axes where 
higher values (i.e. modularity-increasing) are represented as points with higher distance from the center.
For example, the higher the modularity regarding the PEP (proportion of electable periods, see Table \ref{table_categories})
parameter, the higher the distance to the center of each chart. The resulting shape (delineated using 
red lines) represents geometrically the degree of modularity of each educational system taking all 
the parameters into account. Ideally, the larger this shape, the more modular that particular educational 
system is according to our categories. Note that we use a minimum of one unit for each axis to improve 
readability. According to this representation, the most modular educational system 
is definitely Finland (Figure \ref{fig:fin}), followed by French-speaking Switzerland in Figure \ref{fig:ch-french} 
(particularly, the Geneva canton), with Lower Saxony and related German federal states (Figure \ref{fig:de-saxony}), Bremen (Figure \ref{fig:de-br})
and Rhineland-Palatinate (Figure \ref{fig:de-rp}) thereafter. Finland scores the maximum scores in all 
categories investigated. 
In general, highly-modular sytems have in common a high degree of electable periods 
(medium in case of Rhineland-Palatinate) and individually-oriented election systems, with frequent class-level
overlaps. Interestingly, they also share a election-first approach to scheduling. Other systems like Austria 
(Figure \ref{fig:at}) follow a more conservative approach with previously determined slots in the timetable 
for electives and a smaller proportion of electable courses. It can also be remarked that during our 
interview, expert E1 mentioned not believing that a election-first approach could be feasible, given the 
sheer amount of different possibilities to consider. However, such an approach has been practiced successfully since 
at least the 90s in other parts of Europe, notably in Germany. We therefore need to stress here the important role 
of automated timetabling tools in delivering support for complex modular educational systems. It is also 
important for experts to know what automated timetabling tools can do in this area.

\begin{landscape}
\begin{figure}[bt]
    \captionsetup[subfigure]{justification=centering}
\begin{tabular}{ccccc}
    \centering

    \begin{subfigure}[b]{0.25\columnwidth}
 \centering

\begin{tikzpicture}[scale=0.5]
  \path (0:0cm) coordinate (O); 

  \foreach \X in {1,...,\D}{
    \draw (\X*\A:0) -- (\X*\A:\R);
  }
  \foreach \Y in {0,...,\U}{
    \foreach \X in {1,...,\D}{
      \path (\X*\A:\Y*\R/\U) coordinate (D\X-\Y);
      \fill (D\X-\Y) circle (1pt);
    }
    \draw [opacity=0.3] (0:\Y*\R/\U) \foreach \X in {1,...,\D}{
        -- (\X*\A:\Y*\R/\U)
    } -- cycle;
  }
 \foreach \Y in {0,...,\U}{
  \foreach \X in {1,...,\D}{
  \foreach \Z in {0,...,9}{
\path  (\X*\A:\Y.\Z*\R/\U) coordinate (D\X-\Y-\Z);
}
}
}
  \path (1*\A:\L) node (L1) {\tiny ES};
  \path (2*\A:\L) node (L2) {\tiny PEP};
  \path (3*\A:\L) node (L3) {\tiny SC};
  \path (4*\A:\L+30) node (L4) {\tiny CLO};
  \path (5*\A:\L) node (L5) {\tiny SW};
  \path (6*\A:\L) node (L6) {\tiny T};
  \path (7*\A:\L) node (L7) {\tiny ML};
  \path (8*\A:\L+20) node (L8) {\tiny SE};
  \draw [color=red,line width=1.5pt,opacity=0.5]

        (D1-10-0) --
        (D2-1-0) --
        (D3-1-0) --
        (D4-10-0) --
        (D5-1-0) --
        (D6-2) --
        (D7-1-0) --
        (D8-1-0)  -- cycle;
    \end{tikzpicture}
    \caption{Austria}
    \label{fig:at}
    
  \end{subfigure}
   &
    \begin{subfigure}[b]{0.25\columnwidth}
 \centering

\begin{tikzpicture}[scale=0.5]
  \path (0:0cm) coordinate (O); 

  \foreach \X in {1,...,\D}{
    \draw (\X*\A:0) -- (\X*\A:\R);
  }
  \foreach \Y in {0,...,\U}{
    \foreach \X in {1,...,\D}{
      \path (\X*\A:\Y*\R/\U) coordinate (D\X-\Y);
      \fill (D\X-\Y) circle (1pt);
    }
    \draw [opacity=0.3] (0:\Y*\R/\U) \foreach \X in {1,...,\D}{
        -- (\X*\A:\Y*\R/\U)
    } -- cycle;
  }
 \foreach \Y in {0,...,\U}{
  \foreach \X in {1,...,\D}{
  \foreach \Z in {0,...,9}{
\path  (\X*\A:\Y.\Z*\R/\U) coordinate (D\X-\Y-\Z);
}
}
}
  \path (1*\A:\L) node (L1) {\tiny ES};
  \path (2*\A:\L) node (L2) {\tiny PEP};
  \path (3*\A:\L) node (L3) {\tiny SC};
  \path (4*\A:\L+30) node (L4) {\tiny CLO};
  \path (5*\A:\L) node (L5) {\tiny SW};
  \path (6*\A:\L) node (L6) {\tiny T};
  \path (7*\A:\L) node (L7) {\tiny ML};
  \path (8*\A:\L+20) node (L8) {\tiny SE};
  \draw [color=red,line width=1.5pt,opacity=0.5]

        (D1-10-0) --
        (D2-10-0) --
        (D3-10-0) --
        (D4-10-0) --
        (D5-10-0) --
        (D6-10) --
        (D7-10-0) --
        (D8-10-0)  -- cycle;
    \end{tikzpicture}
    \caption{Finland}
    \label{fig:fin}
    
  \end{subfigure}
   & 
    \begin{subfigure}[b]{0.25\columnwidth}
 \centering

\begin{tikzpicture}[scale=0.5]
  \path (0:0cm) coordinate (O); 

  \foreach \X in {1,...,\D}{
    \draw (\X*\A:0) -- (\X*\A:\R);
  }
  \foreach \Y in {0,...,\U}{
    \foreach \X in {1,...,\D}{
      \path (\X*\A:\Y*\R/\U) coordinate (D\X-\Y);
      \fill (D\X-\Y) circle (1pt);
    }
    \draw [opacity=0.3] (0:\Y*\R/\U) \foreach \X in {1,...,\D}{
        -- (\X*\A:\Y*\R/\U)
    } -- cycle;
  }
 \foreach \Y in {0,...,\U}{
  \foreach \X in {1,...,\D}{
  \foreach \Z in {0,...,9}{
\path  (\X*\A:\Y.\Z*\R/\U) coordinate (D\X-\Y-\Z);
}
}
}
  \path (1*\A:\L) node (L1) {\tiny ES};
  \path (2*\A:\L) node (L2) {\tiny PEP};
  \path (3*\A:\L) node (L3) {\tiny SC};
  \path (4*\A:\L+30) node (L4) {\tiny CLO};
  \path (5*\A:\L) node (L5) {\tiny SW};
  \path (6*\A:\L) node (L6) {\tiny T};
  \path (7*\A:\L) node (L7) {\tiny ML};
  \path (8*\A:\L+20) node (L8) {\tiny SE};
  \draw [color=red,line width=1.5pt,opacity=0.5]

        (D1-10-0) --
        (D2-10-0) --
        (D3-10-0) --
        (D4-1-0) --
        (D5-10-0) --
        (D6-2) --
        (D7-1-0) --
        (D8-1-0)  -- cycle;
    \end{tikzpicture}
    \caption{Baden-W\"urttemberg (AGY)}
    \label{fig:de-bw-agy}
    
  \end{subfigure}
   &
    \begin{subfigure}[b]{0.25\columnwidth}
 \centering

\begin{tikzpicture}[scale=0.5]
  \path (0:0cm) coordinate (O); 

  \foreach \X in {1,...,\D}{
    \draw (\X*\A:0) -- (\X*\A:\R);
  }
  \foreach \Y in {0,...,\U}{
    \foreach \X in {1,...,\D}{
      \path (\X*\A:\Y*\R/\U) coordinate (D\X-\Y);
      \fill (D\X-\Y) circle (1pt);
    }
    \draw [opacity=0.3] (0:\Y*\R/\U) \foreach \X in {1,...,\D}{
        -- (\X*\A:\Y*\R/\U)
    } -- cycle;
  }
 \foreach \Y in {0,...,\U}{
  \foreach \X in {1,...,\D}{
  \foreach \Z in {0,...,9}{
\path  (\X*\A:\Y.\Z*\R/\U) coordinate (D\X-\Y-\Z);
}
}
}
  \path (1*\A:\L) node (L1) {\tiny ES};
  \path (2*\A:\L) node (L2) {\tiny PEP};
  \path (3*\A:\L) node (L3) {\tiny SC};
  \path (4*\A:\L+30) node (L4) {\tiny CLO};
  \path (5*\A:\L) node (L5) {\tiny SW};
  \path (6*\A:\L) node (L6) {\tiny T};
  \path (7*\A:\L) node (L7) {\tiny ML};
  \path (8*\A:\L+20) node (L8) {\tiny SE};
  \draw [color=red,line width=1.5pt,opacity=0.5]

        (D1-5-0) --
        (D2-5-0) --
        (D3-10-0) --
        (D4-1-0) --
        (D5-10-0) --
        (D6-2) --
        (D7-1-0) --
        (D8-1-0)  -- cycle;
    \end{tikzpicture}
    \caption{Baden-W\"urttemberg (BGY)}
    \label{fig:de-bw-bgy}
    
  \end{subfigure}
   & \\
    \begin{subfigure}[b]{0.25\columnwidth}
 \centering

\begin{tikzpicture}[scale=0.5]
  \path (0:0cm) coordinate (O); 

  \foreach \X in {1,...,\D}{
    \draw (\X*\A:0) -- (\X*\A:\R);
  }
  \foreach \Y in {0,...,\U}{
    \foreach \X in {1,...,\D}{
      \path (\X*\A:\Y*\R/\U) coordinate (D\X-\Y);
      \fill (D\X-\Y) circle (1pt);
    }
    \draw [opacity=0.3] (0:\Y*\R/\U) \foreach \X in {1,...,\D}{
        -- (\X*\A:\Y*\R/\U)
    } -- cycle;
  }
 \foreach \Y in {0,...,\U}{
  \foreach \X in {1,...,\D}{
  \foreach \Z in {0,...,9}{
\path  (\X*\A:\Y.\Z*\R/\U) coordinate (D\X-\Y-\Z);
}
}
}
  \path (1*\A:\L) node (L1) {\tiny ES};
  \path (2*\A:\L) node (L2) {\tiny PEP};
  \path (3*\A:\L) node (L3) {\tiny SC};
  \path (4*\A:\L+30) node (L4) {\tiny CLO};
  \path (5*\A:\L) node (L5) {\tiny SW};
  \path (6*\A:\L) node (L6) {\tiny T};
  \path (7*\A:\L) node (L7) {\tiny ML};
  \path (8*\A:\L+20) node (L8) {\tiny SE};
  \draw [color=red,line width=1.5pt,opacity=0.5]

        (D1-7-5) --
        (D2-5-0) --
        (D3-1-0) --
        (D4-10-0) --
        (D5-10-0) --
        (D6-1-0) --
        (D7-10-0) --
        (D8-1-0)  -- cycle;
    \end{tikzpicture}
    \caption{Bavaria (FOS/BOS)}
    \label{fig:de-by-fosbos}
    
  \end{subfigure}
   & 
    \begin{subfigure}[b]{0.25\columnwidth}
 \centering

\begin{tikzpicture}[scale=0.5]
  \path (0:0cm) coordinate (O); 

  \foreach \X in {1,...,\D}{
    \draw (\X*\A:0) -- (\X*\A:\R);
  }
  \foreach \Y in {0,...,\U}{
    \foreach \X in {1,...,\D}{
      \path (\X*\A:\Y*\R/\U) coordinate (D\X-\Y);
      \fill (D\X-\Y) circle (1pt);
    }
    \draw [opacity=0.3] (0:\Y*\R/\U) \foreach \X in {1,...,\D}{
        -- (\X*\A:\Y*\R/\U)
    } -- cycle;
  }
 \foreach \Y in {0,...,\U}{
  \foreach \X in {1,...,\D}{
  \foreach \Z in {0,...,9}{
\path  (\X*\A:\Y.\Z*\R/\U) coordinate (D\X-\Y-\Z);
}
}
}
  \path (1*\A:\L) node (L1) {\tiny ES};
  \path (2*\A:\L) node (L2) {\tiny PEP};
  \path (3*\A:\L) node (L3) {\tiny SC};
  \path (4*\A:\L+30) node (L4) {\tiny CLO};
  \path (5*\A:\L) node (L5) {\tiny SW};
  \path (6*\A:\L) node (L6) {\tiny T};
  \path (7*\A:\L) node (L7) {\tiny ML};
  \path (8*\A:\L+20) node (L8) {\tiny SE};
  \draw [color=red,line width=1.5pt,opacity=0.5]

        (D1-10-0) --
        (D2-5-0) --
        (D3-10-0) --
        (D4-10-0) --
        (D5-10-0) --
        (D6-1-0) --
        (D7-1-0) --
        (D8-1-0)  -- cycle;
    \end{tikzpicture}
    \caption{Rhineland-Palatinate}
    \label{fig:de-rp}
    
  \end{subfigure}
   &
    \begin{subfigure}[b]{0.25\columnwidth}
 \centering

\begin{tikzpicture}[scale=0.5]
  \path (0:0cm) coordinate (O); 

  \foreach \X in {1,...,\D}{
    \draw (\X*\A:0) -- (\X*\A:\R);
  }
  \foreach \Y in {0,...,\U}{
    \foreach \X in {1,...,\D}{
      \path (\X*\A:\Y*\R/\U) coordinate (D\X-\Y);
      \fill (D\X-\Y) circle (1pt);
    }
    \draw [opacity=0.3] (0:\Y*\R/\U) \foreach \X in {1,...,\D}{
        -- (\X*\A:\Y*\R/\U)
    } -- cycle;
  }
 \foreach \Y in {0,...,\U}{
  \foreach \X in {1,...,\D}{
  \foreach \Z in {0,...,9}{
\path  (\X*\A:\Y.\Z*\R/\U) coordinate (D\X-\Y-\Z);
}
}
}
  \path (1*\A:\L) node (L1) {\tiny ES};
  \path (2*\A:\L) node (L2) {\tiny PEP};
  \path (3*\A:\L) node (L3) {\tiny SC};
  \path (4*\A:\L+30) node (L4) {\tiny CLO};
  \path (5*\A:\L) node (L5) {\tiny SW};
  \path (6*\A:\L) node (L6) {\tiny T};
  \path (7*\A:\L) node (L7) {\tiny ML};
  \path (8*\A:\L+20) node (L8) {\tiny SE};
  \draw [color=red,line width=1.5pt,opacity=0.5]

        (D1-5-0) --
        (D2-10-0) --
        (D3-10-0) --
        (D4-10-0) --
        (D5-10-0) --
        (D6-1-0) --
        (D7-1-0) --
        (D8-1-0)  -- cycle;
    \end{tikzpicture}
    \caption{Bremen}
    \label{fig:de-br}
    
  \end{subfigure}
   &
    \begin{subfigure}[b]{0.25\columnwidth}
 \centering

\begin{tikzpicture}[scale=0.5]
  \path (0:0cm) coordinate (O); 

  \foreach \X in {1,...,\D}{
    \draw (\X*\A:0) -- (\X*\A:\R);
  }
  \foreach \Y in {0,...,\U}{
    \foreach \X in {1,...,\D}{
      \path (\X*\A:\Y*\R/\U) coordinate (D\X-\Y);
      \fill (D\X-\Y) circle (1pt);
    }
    \draw [opacity=0.3] (0:\Y*\R/\U) \foreach \X in {1,...,\D}{
        -- (\X*\A:\Y*\R/\U)
    } -- cycle;
  }
 \foreach \Y in {0,...,\U}{
  \foreach \X in {1,...,\D}{
  \foreach \Z in {0,...,9}{
\path  (\X*\A:\Y.\Z*\R/\U) coordinate (D\X-\Y-\Z);
}
}
}
  \path (1*\A:\L) node (L1) {\tiny ES};
  \path (2*\A:\L) node (L2) {\tiny PEP};
  \path (3*\A:\L) node (L3) {\tiny SC};
  \path (4*\A:\L+30) node (L4) {\tiny CLO};
  \path (5*\A:\L) node (L5) {\tiny SW};
  \path (6*\A:\L) node (L6) {\tiny T};
  \path (7*\A:\L) node (L7) {\tiny ML};
  \path (8*\A:\L+20) node (L8) {\tiny SE};
  \draw [color=red,line width=1.5pt,opacity=0.5]

        (D1-10-0) --
        (D2-10-0) --
        (D3-10-0) --
        (D4-10-0) --
        (D5-10-0) --
        (D6-1-0) --
        (D7-1-0) --
        (D8-5-0)  -- cycle;
    \end{tikzpicture}
    \caption{Saxony/Lower Saxony/Saxony-Anhalt}
    \label{fig:de-saxony}
    
  \end{subfigure}
   & \\
    \begin{subfigure}[b]{0.25\columnwidth}
 \centering

\begin{tikzpicture}[scale=0.5]
  \path (0:0cm) coordinate (O); 

  \foreach \X in {1,...,\D}{
    \draw (\X*\A:0) -- (\X*\A:\R);
  }
  \foreach \Y in {0,...,\U}{
    \foreach \X in {1,...,\D}{
      \path (\X*\A:\Y*\R/\U) coordinate (D\X-\Y);
      \fill (D\X-\Y) circle (1pt);
    }
    \draw [opacity=0.3] (0:\Y*\R/\U) \foreach \X in {1,...,\D}{
        -- (\X*\A:\Y*\R/\U)
    } -- cycle;
  }
 \foreach \Y in {0,...,\U}{
  \foreach \X in {1,...,\D}{
  \foreach \Z in {0,...,9}{
\path  (\X*\A:\Y.\Z*\R/\U) coordinate (D\X-\Y-\Z);
}
}
}
  \path (1*\A:\L) node (L1) {\tiny ES};
  \path (2*\A:\L) node (L2) {\tiny PEP};
  \path (3*\A:\L) node (L3) {\tiny SC};
  \path (4*\A:\L+30) node (L4) {\tiny CLO};
  \path (5*\A:\L) node (L5) {\tiny SW};
  \path (6*\A:\L) node (L6) {\tiny T};
  \path (7*\A:\L) node (L7) {\tiny ML};
  \path (8*\A:\L+20) node (L8) {\tiny SE};
  \draw [color=red,line width=1.5pt,opacity=0.5]

        (D1-10-0) --
        (D2-10-0) --
        (D3-1-0) --
        (D4-10-0) --
        (D5-10-0) --
        (D6-1-0) --
        (D7-10-0) --
        (D8-10-0)  -- cycle;
    \end{tikzpicture}
    \caption{Luxembourgh}
    \label{fig:lu}
    
  \end{subfigure}
   & 
    \begin{subfigure}[b]{0.25\columnwidth}
 \centering

\begin{tikzpicture}[scale=0.5]
  \path (0:0cm) coordinate (O); 

  \foreach \X in {1,...,\D}{
    \draw (\X*\A:0) -- (\X*\A:\R);
  }
  \foreach \Y in {0,...,\U}{
    \foreach \X in {1,...,\D}{
      \path (\X*\A:\Y*\R/\U) coordinate (D\X-\Y);
      \fill (D\X-\Y) circle (1pt);
    }
    \draw [opacity=0.3] (0:\Y*\R/\U) \foreach \X in {1,...,\D}{
        -- (\X*\A:\Y*\R/\U)
    } -- cycle;
  }
 \foreach \Y in {0,...,\U}{
  \foreach \X in {1,...,\D}{
  \foreach \Z in {0,...,9}{
\path  (\X*\A:\Y.\Z*\R/\U) coordinate (D\X-\Y-\Z);
}
}
}
  \path (1*\A:\L) node (L1) {\tiny ES};
  \path (2*\A:\L) node (L2) {\tiny PEP};
  \path (3*\A:\L) node (L3) {\tiny SC};
  \path (4*\A:\L+30) node (L4) {\tiny CLO};
  \path (5*\A:\L) node (L5) {\tiny SW};
  \path (6*\A:\L) node (L6) {\tiny T};
  \path (7*\A:\L) node (L7) {\tiny ML};
  \path (8*\A:\L+20) node (L8) {\tiny SE};
  \draw [color=red,line width=1.5pt,opacity=0.5]

        (D1-5-0) --
        (D2-5-0) --
        (D3-1-0) --
        (D4-1-0) --
        (D5-10-0) --
        (D6-1-0) --
        (D7-1-0) --
        (D8-5-0)  -- cycle;
    \end{tikzpicture}
    \caption{Netherlands}
    \label{fig:nl}
    
  \end{subfigure}
   & 
    \begin{subfigure}[b]{0.25\columnwidth}
 \centering

\begin{tikzpicture}[scale=0.5]
  \path (0:0cm) coordinate (O); 

  \foreach \X in {1,...,\D}{
    \draw (\X*\A:0) -- (\X*\A:\R);
  }
  \foreach \Y in {0,...,\U}{
    \foreach \X in {1,...,\D}{
      \path (\X*\A:\Y*\R/\U) coordinate (D\X-\Y);
      \fill (D\X-\Y) circle (1pt);
    }
    \draw [opacity=0.3] (0:\Y*\R/\U) \foreach \X in {1,...,\D}{
        -- (\X*\A:\Y*\R/\U)
    } -- cycle;
  }
 \foreach \Y in {0,...,\U}{
  \foreach \X in {1,...,\D}{
  \foreach \Z in {0,...,9}{
\path  (\X*\A:\Y.\Z*\R/\U) coordinate (D\X-\Y-\Z);
}
}
}
  \path (1*\A:\L) node (L1) {\tiny ES};
  \path (2*\A:\L) node (L2) {\tiny PEP};
  \path (3*\A:\L) node (L3) {\tiny SC};
  \path (4*\A:\L+30) node (L4) {\tiny CLO};
  \path (5*\A:\L) node (L5) {\tiny SW};
  \path (6*\A:\L) node (L6) {\tiny T};
  \path (7*\A:\L) node (L7) {\tiny ML};
  \path (8*\A:\L+20) node (L8) {\tiny SE};
  \draw [color=red,line width=1.5pt,opacity=0.5]

        (D1-10-0) --
        (D2-1-0) --
        (D3-10-0) --
        (D4-1-0) --
        (D5-10-0) --
        (D6-2) --
        (D7-1-0) --
        (D8-1-0)  -- cycle;
    \end{tikzpicture}
    \caption{German-speaking Switzerland}
    \label{fig:ch-german}
    
  \end{subfigure}
   & 
    \begin{subfigure}[b]{0.25\columnwidth}
 \centering

\begin{tikzpicture}[scale=0.5]
  \path (0:0cm) coordinate (O); 

  \foreach \X in {1,...,\D}{
    \draw (\X*\A:0) -- (\X*\A:\R);
  }
  \foreach \Y in {0,...,\U}{
    \foreach \X in {1,...,\D}{
      \path (\X*\A:\Y*\R/\U) coordinate (D\X-\Y);
      \fill (D\X-\Y) circle (1pt);
    }
    \draw [opacity=0.3] (0:\Y*\R/\U) \foreach \X in {1,...,\D}{
        -- (\X*\A:\Y*\R/\U)
    } -- cycle;
  }
 \foreach \Y in {0,...,\U}{
  \foreach \X in {1,...,\D}{
  \foreach \Z in {0,...,9}{
\path  (\X*\A:\Y.\Z*\R/\U) coordinate (D\X-\Y-\Z);
}
}
}
  \path (1*\A:\L) node (L1) {\tiny ES};
  \path (2*\A:\L) node (L2) {\tiny PEP};
  \path (3*\A:\L) node (L3) {\tiny SC};
  \path (4*\A:\L+30) node (L4) {\tiny CLO};
  \path (5*\A:\L) node (L5) {\tiny SW};
  \path (6*\A:\L) node (L6) {\tiny T};
  \path (7*\A:\L) node (L7) {\tiny ML};
  \path (8*\A:\L+20) node (L8) {\tiny SE};
  \draw [color=red,line width=1.5pt,opacity=0.5]

        (D1-10-0) --
        (D2-10-0) --
        (D3-10-0) --
        (D4-10-0) --
        (D5-10-0) --
        (D6-1-0) --
        (D7-10-0) --
        (D8-10-0)  -- cycle;
    \end{tikzpicture}
    \caption{French-speaking Switzerland}
    \label{fig:ch-french}
    
  \end{subfigure}
   & \\ 
\end{tabular}
\caption{Visual comparison using radar charts from the educational systems studied in this paper.}\label{fig:radarcharts}
\label{fig_comparison}
\end{figure}
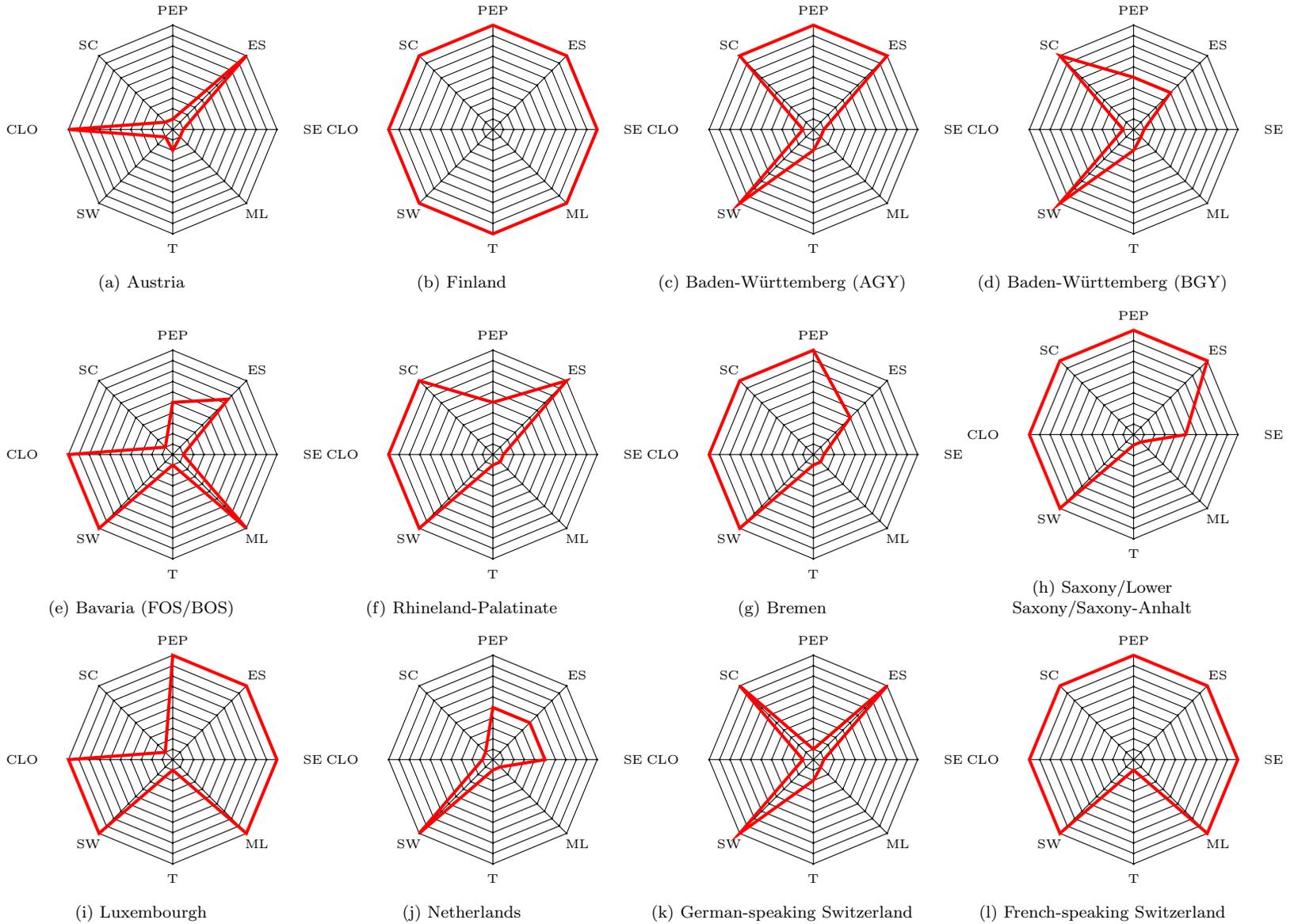
\end{landscape}


\subsection{Consequences for scheduling}

The interviews show that there are various challenges when trying to create a schedule that satisfies all the needs of students, 
teachers and schools. The central problem shared by all modular educational systems studied is how to assign courses to students 
in such a way that students can attend all the courses they have elected. For instance, there might not be enough teachers for 
courses where too many students registered, or courses might need to be canceled when not enough students decide to enroll. In these
cases, it will be necessary to relocate students to other courses. Additionally, course occupation needs to be balanced as well. E.g. 
if there are two advanced math courses that are equivalent (i.e. parallel courses), it should be avoided that the number 
of students that attend each course is very different.

Another aspect that came up frequently in the interviews was the necessity to minimize the number of 
idle periods for students (and, secondarily, for teachers as well). In class-based systems, since students attend lessons 
together, the timetabler can easily ensure that core times are always assigned a lesson, resulting in no idle times 
for the entire class. However, this can be significantly more challenging in modular systems, since every student has their own timetable.
As a result, some timetablers (mainly in Germany) use a two-step method for building the schedule: First, they cluster courses
in such a way that these courses can be scheduled simultaneously (no teacher or student clashes). Second, they maximize the amount 
of students in every cluster, which means that when a cluster is scheduled in a given time slot, no student will have idle time then.
If these clusters can completely fill a set of core time slots, it can be guaranteed that no idle periods are produced. However, 
this task can be extremely challenging and automated tools are needed. In some systems (e.g. the Netherlands), timetablers try to 
avoid class level overlaps to make this problem easier to solve to some extent. Even then, it might be impossible to avoid idle times at all. 
How difficult these problems turn out to be is largely determined by the PEP (proportion of electable periods) and ES (election system)
categories in our framework. In general, the higher the PEP, the more possibilities need to be considered when building the timetable.
The more individualized an election system, the more difficult it is to take all student choices into consideration. 

Lastly, the room situation and teacher availability needs to be considered as well. If too many courses are scheduled at the 
same time that need the same type of room (e.g. chemistry lab, projectors, etc), the timetable might be unfeasible. 
Similarly, teachers working at more 
than one school need to be considered separately since they might not be available at certain days or times. Modular systems with 
school collaborations face this type of problem when building timetables (e.g. in Finland). 

In general, the higher the modularity as holistically estimated by our framework, the more 
difficult building a timetable becomes. The problem of creating a timetable that satisfies a given set of requirements is not new and has been studied for more than 50 years \citep{even_complexity_1975}.
Throughout the years the problem of educational timetabling has evolved into three subproblems that are often treated as independent: High school timetabling, course timetabling and examination timetabling \citep{schaerf_survey_1999}.

In high school timetabling the focus lies in creating a weekly schedule for the various lessons of classes in a way that gives both students and teachers a compact schedule according to their availabilites while also paying attention to room assignments and other organizational constraints.

Course timetabling centers around the assignment of students to classes that are then scheduled in a timetable where each week of the term is scheduled individually.
Again many different organizational constraints like assigning rooms and paying attention to the availabilities of teachers apply but the main focus lies in scheduling the lessons in a way such that every student can attend their desired courses.
This makes it a good fit for creating schedules for universities.

Finally, in examination timetabling the goal is to assign exams to examination periods in a way that every student can take the exams they signed up for.
This problem differs from the others since examinations are usually singular events without patterns that extend over several weeks or months. 
However, many of the constraints from other educational timetabling problems, like room requirements and distributing events, still apply.

From the interviews we have conducted, we can conclude that none of the three mentioned categories completely 
cover the requirements for the regions in which we investigated. 
For instance, course timetabling focuses on the assignment of students to classes, usually not considering other aspects like 
compactness or pedagogical considerations like scheduling lesson in double periods. Both student assignments and pedagogical 
considerations are relevant in all educational systems studied.
The problem definition seems to lie somewhere between high school timetabling and course timetabling. 
However, combining those two disciplines is no easy task since each on its own already is difficult enough to inspire decades of research that still results in non-optimal schedules.
Nevertheless, we want to move the research in a direction that focuses on more modern school systems with modular course choices for students.
To achieve this, this study represents the first step in a series of papers where we will propose new techniques for improving automatic scheduling software.
We hope that developing such automatic timetabling software that supports the requirements outlined in this paper will make it easier for schools to transition to modular education 
which in turn will give students many more opportunities to follow their interests and prepare for their future careers through individualized choices.

\section{Conclusion}\label{conclusion}

In this paper, we presented a selection of modular educational systems in Europe and a novel category system for classifying 
them according to a well-defined set of characteristics. To the best of our knowledge, this is the first effort of this kind 
in the literature to date. This study lays the groundwork for analyizing the needs and requirements of timetabling in different 
regions when modular courses are present, paving the way for a detailed investigation into how to automate this requirements 
to build specially tailored tools that support realizing the promise of modular education in a way that improves its quality 
for all stakeholders involved, including students, teachers, school administration and policy makers.

For this investigation, we had access to interview partners covering a wide range of educational systems. However, the studied regions
by far don't cover the whole of secondary education in Europe. We expect the selection presented in this paper to be representative 
of the main trends in modern modular education, but prominent examples (like the United Kingdom) are missing. We plan to fill this gap 
in future work. 

Additionally, automatic timetabling methods that comply with the requirements investigated in this paper will be developed to 
support educational systems across Europe to provide and extend modular education at the secondary level.

\section{Acknowledgements}

The authors acknowledge financial support from the Agency for Research Promotion of Lower Austria (GFF) under Project 
\enquote{Course scheduling in modular educational systems}.



\bibliographystyle{elsarticle-harv} 
\bibliography{bibliography.bib}




\end{document}